\pdfoutput=1
\documentclass[fleqn,useAMS,usenatbib]{mn2e}
\usepackage{./aas_macros}

\usepackage{amsmath}
\usepackage{txfonts}
\usepackage{natbib}
\usepackage[dvips]{graphicx}

\usepackage{hyperref}
\usepackage[usenames,dvipsnames]{color}
\definecolor{menublue}{rgb}{0.0,0.0,0.5}
\definecolor{citegreen}{rgb}{0.0,1.0,0.0}
\definecolor{urlred}{rgb}{1.0,0.0,0.0}
\hypersetup{bookmarksopen,pdfstartview={FitH},colorlinks=true, breaklinks=true,menucolor=menublue,urlcolor=urlred,citecolor=citegreen,linkcolor=blue}
\usepackage[all]{hypcap}

\usepackage{graphicx}
\usepackage{caption}
\usepackage{subfigure}

\def\del#1{{}}

\newcommand{\ltsima}{$\; \buildrel < \over \sim \;$}
\newcommand{\lsim}{\lower.5ex\hbox{\ltsima}}
\newcommand{\gtsima}{$\; \buildrel > \over \sim \;$}
\newcommand{\gsim}{\lower.5ex\hbox{\gtsima}}

\newcommand{\chip}{{\chi^\prime}}
\newcommand{\chipp}{{\chi^{\prime\prime}}}

\title[CMB-lensing with Born-corrections]
{Born-corrections to weak lensing of the cosmic microwave background temperature and polarisation anisotropies}
\author[Steffen Hagstotz,  Bj{\"o}rn Malte Sch{\"a}fer and Philipp M. Merkel]
{Steffen Hagstotz\thanks{e-mail: steffen.hagstotz@gmail.de}$^{1,2,3}$, Bj{\"o}rn Malte Sch\"afer$^1$ and Philipp M. Merkel$^4$\\
$^1$Astronomisches Recheninstitut, Zentrum f{\"u}r Astronomie der Universit{\"a}t Heidelberg, Philosophenweg 12, 69120 Heidelberg, Germany\\
$^2$Universit\"ats-Sternwarte, Fakult\"at f\"ur Physik, Ludwig-Maximilians Universit\"at M\"unchen, Scheinerstr. 1, 81679 M\"unchen, Germany\\
$^3$Excellence Cluster Universe, Boltzmannstr. 2, 85748 Garching, Germany\\
$^4$Institut f{\"u}r Theoretische Astrophysik, Zentrum f{\"u}r Astronomie der Universit{\"a}t Heidelberg, Philosophenweg 12, 69120 Heidelberg, Germany}

\begin{document}
\pagerange{\pageref{firstpage}--\pageref{lastpage}}
\pubyear{2014}
\maketitle
\label{firstpage}

% --- abstract --- %
\begin{abstract}
Many weak lensing calculations make use of the Born approximation where the light ray is approximated by a straight path. We examine the effect of Born-corrections for lensing of the cosmic microwave background in an analytical approach by taking perturbative corrections to the geodesic into account. The resulting extra power in the lensing potential spectrum is comparable to the power generated by nonlinear structure formation and affects especially the polarisation spectra, leading to relative changes of the order of one per cent for the $E$-mode spectrum and up to 10 per cent on all scales to the $B$-mode spectrum. In contrast, there is only little change of spectra involving the CMB temperature. Additionally, the corrections excite one more degree of freedom resulting in a deflection component which can not be described as a gradient of the lensing potential as it is related to image rotation in lens-lens coupling. We estimate the magnitude of this effect on the CMB-spectra and find it to be negligible.
\end{abstract}

% --- keywords --- %

% --- section: introduction --- %
\section{Introduction}\label{sec_introduction}
Weak gravitational lensing by the cosmic large-scale structure is a primary tool for investigating cosmological models through their influence on the growth of structure. In weak lensing, one observes either correlated shape changes of distant galaxies and traces them back to correlated distortion along neighbouring lines of sight, or the change in cosmic microwave background fluctuations due to correlated deflection. Both techniques yield statistically significant signals and have been used experimentally to constrain cosmological models, e.g. \citet{2007PhRvD..76d3510S, 2012ApJ...756..142V, planckxvii, 2014MNRAS.441.2725F}. Weak lensing derives its attractiveness from the linear relationship between the density field and the weak lensing observables, at least to a very good approximation, and due to the dominance of dark matter one deals with comparatively simple physics in structure formation.

In the theory of gravitational lensing one solves the implicit lensing equation in a perturbative expansion, which gives rise to Born-corrections and in the lensing Jacobian to lens-lens coupling effects. These effects have been investigated in the context of weak cosmic shear \citep{Cooray/Hu, Krause} and weak cosmic flexion \citep{Flexion}, where they generate small corrections in the shear spectra. Transferring these results to CMB-lensing one finds two competing arguments: Firstly, weak lensing of the CMB takes place at much larger redshifts due to the lensing efficiency function where the cosmic matter field is less developed and has smaller amplitudes, generating a weaker second order effect. Secondly, Born-corrections and lens-lens couplings are cumulative effects and the much larger line of sight integrations should enhance them: The relative magnitude of these two arguments is the primary motivation for this work.

In the case of CMB-lensing only three main effects beyond linar physics influence the lensing signal: $(i)$ nonlinear structure growth on small scales, which generates an increase in the variance of the lensing deflection field, $(ii)$ non-Gaussian statistics of the underlying density field, and $(iii)$ second order lensing effects such as Born-corrections and lens-lens coupling. The last group is usually neglected with reference to cosmic shear, where they have been shown to be small, but as we will demonstrate in this paper this does not necessarily need to be the case. In contrast, nonlinear structures certainly affect the lensing signal, even at the large redshifts of CMB-lensing, and are modelled by nonlinear extentions to the dark matter spectrum. The validity of this approach is verified in numerical simulations, where only on small scales below an arcminute lensing by individual haloes starts to matter \citep{2005MNRAS.356..829H}. The effect of nonlinear structure is primarily the increase of variance of the deflection field on small scales, while non-Gaussian features do not have a strong impact on the weak lensing signal, which has been shown by a higher-order expansion of the characteristic function of the lensing deflection distribution \citep{PhilippGauss}.

Numerical tools for investigating CMB-lensing rely as well on the Born-approximation: Boltzmann-codes such as \textsc{camb}\footnote{http://camb.info/} or \textsc{class}\footnote{http://class-code.net/} for predicting CMB temperature and polarisation anisotropy spectra use a Limber-projected, Born-approximated deflection angle spectrum for analytical derivations of lensed CMB-spectra. Also \textsc{lenspix}\footnote{http://cosmologist.info/lenspix/} for computing lensed realisations of the cosmic microwave background determines the deflection angle field from the Born-approximated lensing potential.

This paper is organised as follows: We present corrections to the first-order weak lensing statistics in Sect.~\ref{sect_lensing}. We apply them to lensing of the cosmic microwave background and discuss the effect of higher-order effects in gravitational lensing in Sect.~\ref{sect_cmb}. The results are summarised and discussed in Sect.~\ref{sect_summary}. 

The reference cosmological model used is a spatially flat $w$CDM cosmology with Gaussian adiabatic initial perturbations in the cold dark matter. The dark energy component is assumed to be spatially homogeneous and to be described with a constant equation of state parameter $w$. The specific parameter choices are $\Omega_\mathrm{m} = 0.3$, $n_\mathrm{s} = 1$, $\sigma_8 = 0.82$, $\Omega_\mathrm{b}=0.045$ and $H_0=100\: h\:\mathrm{km}\:\mathrm{s}^{-1}\:\mathrm{Mpc}^{-1}$, with $h=0.7$ and $w=-0.9$.

% --- section: weak gravitational lensing --- %
\section{Weak Gravitational Lensing}\label{sect_lensing}

% ---  --- %
\subsection{Lens equation}
We will describe gravitational lensing by expressing the comoving distance $\bmath x (\btheta, \chi)$ between the deflected light ray and a fiducial unperturbed ray which enclose the angle $\btheta$ at the observer. This separation evolves in the presence of gravitational fields according to the lens equation \citep{bartelmann/schneider, Bartelmann}:
\begin{equation}\label{eq_lens_x}
\bmath x (\btheta, \chi) =\btheta \chi - 2 \int_0^\chi \mathrm d \chi^\prime \: (\chi - \chi ^\prime) \:  \nabla_\perp \phi(\btheta, \chi^\prime) \: ,
\end{equation}
where we neglected any anisotropic stress and used the dimensionless Newtonian potential $\phi = \Phi/c^2$. A source with the true angular position $\btheta$ will therefore be observed at the lensed position $\bbeta = \bmath{x} / \chi$, which becomes
\begin{equation}
\beta_i (\btheta, \chi^\prime) = \theta_i - 2 \int_0^\chi \mathrm d \chi^\prime \: \frac{\chi - \chi ^\prime}{\chi} \:  \phi_i(\btheta, \chi) = \theta_i - \psi_i
\end{equation}
where the derivative perpendicular to the line of sight is written as $\upartial_i \phi = \phi_i$ and we may as well express the angles as gradients of the projected lensing potential $\psi$:
\begin{equation}
\psi \equiv 2 \int \mathrm d \chi^\prime \: \frac{\chi - \chi^\prime}{\chi} \: \phi(\btheta, \chi^\prime) \: .
\end{equation}
The linearised mapping between source and image planes is then expressed by the Jacobian matrix $\mathcal{A} \equiv \upartial \bbeta / \upartial \btheta$:
\begin{equation}
\mathcal{A}_{ij} \equiv \frac{\upartial \beta_i}{\upartial \theta_j} =  \begin{pmatrix} 1-\kappa - \gamma_1 & -\gamma_2 - \omega \\ -\gamma_2 + \omega & 1- \kappa + \gamma_1 \end{pmatrix}
\end{equation}
with the convergence $\kappa$, the shear components $\gamma_i$ and the antisymmetric component $\omega$ describing an infinitesimal rotation \citep{2006MNRAS.367.1543P}. These coefficients also define the decomposition into Pauli matrices $\sigma_\alpha$ which, including $\sigma_0 \equiv \mathrm{id}(2)$, form a basis set for the $2 \times 2$ matrices:
\begin{equation}
\mathcal{A} = \sum_{\alpha=0}^{3} a_\alpha \sigma_\alpha \equiv (1-\kappa) \sigma_0 - \gamma_1 \sigma_3 - \gamma_2 \sigma_1 + \mathrm{i} \omega \sigma_2 \: .
\end{equation}
Because the Jacobian is composed of interchangable partial derivatives to first order, it is symmetric and the $\omega$-component vanishes: Therefore, only three of the four possible degrees of freedom are excited. Due to the properties of the Pauli matrices $\sigma_\alpha^2 = \mathrm{id}(2)$, $\mathrm{tr}(\sigma_i) = 0$ for $i = 1,2,3$ and $\sigma_\alpha \sigma_\beta = \delta_{\alpha \beta}\sigma_0 + i \epsilon_{\alpha \beta \gamma} \sigma_\gamma$ the coefficients can be recovered via $a_\alpha = \frac{1}{2} \mathrm{tr}(\mathcal{A} \sigma_\alpha)$. Note that $\kappa$, $\gamma_1$ and $\gamma_2$ can be written as second derivatives of the underlying lensing potential
\begin{equation}
\kappa = \frac{1}{2} ( \upartial_1^2 + \upartial_2^2) \: \psi,\quad
\gamma_1 = \frac{1}{2} (\upartial_1^2 - \upartial_2^2 ) \: \psi,\quad
\gamma_2 = \vphantom{\frac{1}{2}} \upartial_1 \upartial_2 \psi,
\end{equation}
whereas for the rotational component $\omega$ this is not the case. Therefore we introduce the auxiliary curl potential $\Omega$ such that the deflection angle may be expressed as a combination of a gradient contribution and a curl contribution \citep{Hirata}
\begin{equation}
\bbeta = \btheta - \nabla \psi - \nabla \times \Omega \: ,
\end{equation}
where we defined the two-dimensional curl $(\nabla \times \Omega)_i = \epsilon_{ij} \upartial_j \Omega$. With this definition, the scalar quantities $\kappa$ and $\omega$ can be obtained by a further derivative from the deflection angle $\delta \btheta = \bbeta - \btheta$:
\begin{equation}\label{eq_kappa_omega}
\kappa = \frac{1}{2} \nabla \cdot \bdelta \btheta = \frac{1}{2} \Delta \psi \: , \hspace{15pt} \omega = \frac{1}{2} \nabla \times \delta \btheta = \frac{1}{2} \Delta \Omega \: .
\end{equation}
In this way we can identify the primary contribution to the lensing deflection which originates as a gradient from the scalar lensing deflection, and a secondary contribution which corresponds to a rotation. 

The deformation part of the Jacobian can be isolated by subtracting the zeroth-order identical mapping and defining the tensor $\Psi_{ij} = \delta_{ij} - \mathcal{A}_{ij}$. We then arrive at the linearised lens equation in terms of $\Psi$:
\begin{equation}\label{eq_lens_psi}
\Psi_{ij}(\bmath{x}, \chi) = 2 \int_0^{\chi} \mathrm{d} \chi^\prime \: W(\chi^\prime, \chi) \: \phi_{mj} (\bmath{x}, \chi) \left[ \delta_{mj} +  \Psi_{mj}(\bmath{x}, \chi^\prime) \right] \:,
\end{equation}
where we introduced the geometric lensing efficiency function
\begin{equation}
W(\chi^\prime, \chi) = \begin{cases}
	(\chi-\chi^\prime)\dfrac{\chi^\prime}{\chi}  & \text{if } \chi^\prime \leq \chi \\
	0					& \text{else}
\end{cases}
\end{equation}
and summation over repeated transversal indices is implied. Note that eq.~\ref{eq_lens_psi} is implicit in both the deformation tensor and the potential along the full geodesic $\phi(\bmath x, \chi)$, where $\bmath{x}(\btheta, \chi)$ again depends on the potential via eq.~\ref{eq_lens_x}. By assuming small initial deformations $\Psi^{(0)} \approx 0$ (neglection of lens-lens coupling) and evaluating the potential along the unperturbed geodesic $\bmath{x}^{(0)} \approx \btheta \chi$ (the Born approximation) one arrives at the well-known first-order result
\begin{equation}\label{eq_psi_first_order}
\Psi^{(1)}_{ij}(\bmath x_0, \chi) = 2 \int_0^\chi \mathrm d \chi^\prime \: W(\chi^\prime, \chi) \: \phi_{ij}(\bmath x_0, \chi^\prime) \: .
\end{equation}
In the literature the use of the Born approximation generally implies neglecting lens-lens coupling as well since they appear at the same order in perturbation theory. We will follow this convention unless explicitly stated otherwise.

Note that the perturbed geodesics change the distance-redshift relation because photon paths are longer taking these corrections into account. This will lead to a distortion of surfaces of equal redshift in real space, but in our Newtonian framework we will neglect this effect. However, see \cite{2014arXiv1405.7860C} for a detailed relativistic treatment.

% ---  --- %
\subsection{Born-corrections}\label{subsect_born}
Corrections to the first-order result have previously been discussed in the literature in the context of cosmic shear surveys both analytically \citep{Cooray/Hu, Shapiro, Krause} and numerically in ray tracing simulations \citep{2000ApJ...530..547J, 2009A&A...499...31H, Li:2010cm, MNR:MNR18754, 2013MNRAS.435..115B} where they were found to be negligible up to very small scales. We will briefly present the correction terms arising when dropping the Born approximation before applying them to CMB lensing.

The first step is to expand the potential $\phi$ around the unperturbed geodesic $\bmath{x}_0$:
\begin{equation}
\phi(\bmath{x}_0 + \delta \bmath x) \approx \phi(\bmath{x}_0) +  \phi_a(\bmath{x}_0) \: \delta x_a + \frac{1}{2} \phi_{ab}(\bmath{x}_0) \: \delta x_a \delta x_b + \dots
\end{equation}
Because we are considering perturbative corrections to the power spectrum $\propto \phi^4$, all terms up to third order in the potential must be kept here because they couple to first order contributions in the correlator later on. In addition to the potential itself, the deflection from the straight path from eq.~\ref{eq_lens_x} depends on the fully deflected light ray:
\begin{equation}
\delta x_a = - 2 \int_0^\chi \mathrm{d} \chi^\prime \: \frac{\chi - \chi^\prime}{\chi} \chi^\prime \: \phi_a( \bmath x, \chi^\prime ) \: .
\end{equation}
Hence we also expand it to get
\begin{equation}
\begin{split}
\delta x^{(1)}_a &= - 2 \int_0^\chi \mathrm{d} \chi^\prime \: \frac{\chi - \chi^\prime}{\chi} \chi^\prime \: \phi_a(\bmath x_0, \chi^\prime ) \\
\delta x^{(2)}_a &=  - 2 \int_0^\chi \mathrm{d} \chi^\prime \: \frac{\chi - \chi^\prime}{\chi} \chi^\prime \: \phi_{ab}(\bmath x_0, \chi^\prime ) \: \delta x_b^{(1)}(\bmath x_0, \chi^\prime) \: ,
\end{split}
\end{equation}
which leads to the complete expression for the potential along the deflected path up to third order:
\begin{equation}
\begin{split}
\phi(\bmath{x}) &\approx \phi(\bmath{x}_0) + \phi_a(\bmath{x}_0) \left( \delta x_a^{(1)} + \delta x_a^{(2)} \right) + \frac{1}{2}  \phi_{ab}(\bmath{x}_0) \: \delta x_a^{(1)} \delta x_b^{(1)} \\
\end{split}
\end{equation}

By iteratively inserting the result for the deformation tensor into the full expression given in eq.~\ref{eq_lens_psi}, $\Psi_{mj} \approx \Psi_{mj}^{(1)}$, one can obtain corrections up to third order in $\phi$:
\begin{equation}\label{eq_corrections}
\begin{split}
 \Psi_{ij} ^{(1)} = & \: 2 \int_0^{\chi} \mathrm{d} \chi^\prime \: W(\chi^\prime, \chi) \: \phi_{ij} (\bmath{x}_0, \chi^\prime) \\ \medskip
 \Psi_{ij} ^{(2)} = & \: 2 \int_0^{\chi} \mathrm{d} \chi^\prime \: W(\chi^\prime, \chi) \: \Big[ \phi_{ij} + \phi_{im} \Psi_{mj}^{(1)} + \phi_{ijk} \delta x_k^{(1)} \Big] \\ \medskip
 \Psi_{ij} ^{(3)} = & \: 2 \int_0^{\chi} \mathrm{d} \chi^\prime \: W(\chi^\prime, \chi) \: \Big[ \phi_{ij} + \phi_{ijk} \left(\delta x_k^{(1)} + \delta x_k^{(2)} \right) + \\ & 
\hphantom{2 \int_0^{\chi} \mathrm{d} \chi^\prime \: W(\chi^\prime, \chi)} 
\frac{1}{2} \phi_{ijkm} \delta x_k^{(1)} \delta x_m^{(1)} + \phi_{im} \Psi_{mj}^{(2)} \Big] \: .
\end{split}
\end{equation}

These expressions are analogous to the Born series in quantum scattering theory and similar diagrammatic representations of the arising terms exist \citep{Flexion}, but we will just be concerned with the final result in this paper.

The various contributions can all physically be attributed to the coupling of lenses at different redshifts of the form $\phi_{im} \Psi_{mj}$, non-local Born-corrections of the form $\phi_{ijk} \delta x_k$ or a combination of both effects at third order. To distinguish the correction terms, we will express the deflection as a projection of various source fields as done in \citet{Cooray/Hu}
\begin{equation}
\Psi_{ij} = 2 \int_0^\chi \mathrm d \chi^\prime \: W(\chi^\prime, \chi) \: \sum_\alpha S^{(\alpha)}_{ij}(\chi^\prime) \: ,
\end{equation}
e.g. with the first order source term $S^{(1)}_{ij} =  \phi_{ij}$. Clearly, the correction terms depend on the correlation of lensing structure along the line of sight: From a physical point of view, the distance from the fiducial ray and the deformation of the light bundle depend on the integrated lensing up to that point, such that  the tidal fields in Fourier space get linked by convolution. All resulting source terms are listed in the Appendix \ref{appendix_born}.

% ---  --- %
\subsection{Deflection Angle Spectra}\label{subsect_spectra}
We will now discuss the statistical properties of the corrections and how they influence the correlation functions of lensing observables. Statistical isotropy requires the power spectrum of the source terms to be of the form
\begin{equation}
\langle S_{ab}(\bmath l) \: S_{ij}^\ast(\bmath l^\prime) \rangle = (2 \upi)^2 \delta_\mathrm{D}(\bmath l - \bmath l^\prime) C_{abij}(\ell) \: ,
\end{equation}
where the source terms contain products of the potential $\phi$ which turn into convolutions in Fourier space. Note that we only consider contributions up to $\phi^4$, so third-order terms only couple to the first-order result whereas under the assumption of Gaussianity all terms $\propto \phi^3$ vanish. The full correlator $C^\Psi_{abij}$ is then recovered as a sum over all source term contributions. The two-point functions of the observables can be extracted via
\begin{equation}
\langle a_\alpha \: a_\beta^\ast \rangle = \frac{1}{4} \langle \mathrm{tr}(\Psi \sigma_\alpha) \: \mathrm{tr}(\Psi \sigma_\beta)^\ast \rangle = \frac{1}{4} C_{abij}^\Psi (\ell) \: \sigma_{\alpha, ab} \sigma_{\beta, ij}^\ast \: . 
\end{equation}
The potential power spectra in the flat-sky limit are given by the Limber equation \citep{Kaiser, Krause, Hirata}:
\begin{equation}
\langle \phi(\bmath{l}, \chi) \: \phi^\ast(\bmath{l}^\prime, \chi^\prime) \rangle = \frac{(2 \upi)^2}{\chi^2} \delta_\mathrm{D}(\bmath{l}-\bmath{l}^\prime) \: \delta_\mathrm{D} (\chi - \chi^\prime) \: P_\phi(k=\ell/\chi)
\end{equation}
and an additional projection yields the angular 2D spectra:
\begin{equation}
C_\ell = \int^{\chi_s}_0 \frac{\mathrm{d} \chi}{\chi^2} \: W^2(\chi, \chi_s) \: P_\phi(k=\ell/\chi) \: .
\end{equation}
All corrections to the power spectra following from eq.~\ref{eq_corrections} are then expressed as line of sight integrations over convolutions of power spectra. The corrections to cosmic shear observables $\kappa$ and $\gamma$ have been discussed by \cite{Cooray/Hu} and \cite{Krause} before, but here we are ultimately interested in CMB lensing. For this purpose, we only need the power spectra of the potentials $C^{\psi \psi}$ and $C^{\Omega \Omega}$ which are related to the $\kappa$ and $\omega$ spectra in Fourier space by eq.~\ref{eq_kappa_omega}:
\begin{equation}
C^{\psi \psi}_\ell= \frac{4}{\ell^4} C^{\kappa \kappa}_\ell \: , \hspace{15pt} C^{\Omega \Omega} _\ell = \frac{4}{\ell^4} C^{\omega \omega}_\ell \: .
\end{equation}
The effect of the image rotation introduced by the curl potential $\Omega$ will be discussed separately in Sect. \ref{sect_curl}. Both spectra for the application in CMB lensing are shown in Fig.~\ref{fig_cpsi_spectrum}, where all line of sight integrals were carried out to the surface of last scattering at $z \approx 1089$.

% ---  --- %
\subsection{Gaussianity}\label{sect_gauss}
The derivation above assumes Gaussian statistics of the underlying lensing potential, so the four-point functions $\langle \phi^4 \rangle$ are completely described by the various two-point correlators via the Wick theorem. Note that the higher moments in the lensing potential $\psi$ are suppressed compared to the density field itself by the central limit theorem because of averaging along the line of sight. Especially in CMB lensing, this approximation is valid up to very small scales as demonstrated by \cite{PhilippGauss}. Non-Gaussianities are small to begin with because the density field is probed at early times, and the line of sight is long compared to cosmic shear surveys.

However, the higher variance of the lensing angle caused by nonlinear structures does influence the result. This is accounted for by using the nonlinear transfer functions provided by \cite{Smith} while still assuming a Gaussian probability distribution of the lensing angles.

The Born-corrections discussed here are treated in the same way. Even though mode coupling is introduced by higher order corrections which will lead to non-Gaussianity in the deflection angle statistics, the dominant effect is the additional amplitude at small scales of the $C^{\psi \psi}_\ell$ power spectrum in Fig.~\ref{fig_cpsi_spectrum}. Thus no higher statistical measures beyond the power spectrum are taken into account for the application to CMB lensing.

% --- section: cmb lensing --- %
\section{CMB Lensing}\label{sect_cmb}

% ---  ---%
\subsection{Lensing potential}

\begin{figure}
\begin{center}
\resizebox{0.95\hsize}{!}{\includegraphics{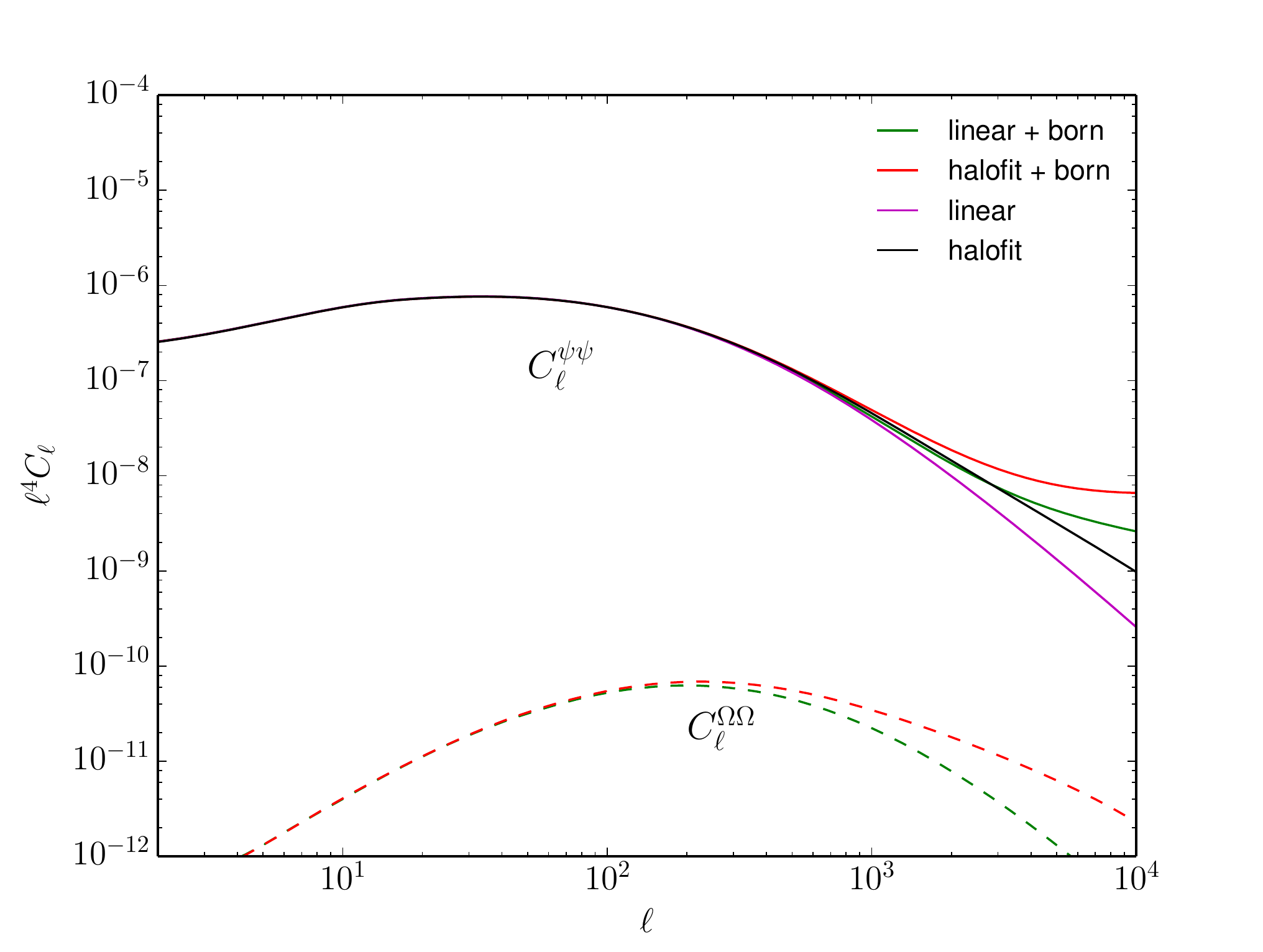}}
\end{center}
\caption{Deflection angle power spectra $C^{\psi \psi}_\ell$ and $C^{\Omega \Omega}_\ell$ including nonlinear structure growth via \textsc{halofit} and including all second order corrections. Shown are the first order result with linear underlying CDM spectrum (violet), nonlinear CDM spectrum (black) and including second order effects in combination with linear (green) and nonlinear power (red). The dashed $C^\Omega$ spectrum only appears if second order corrections are taken into account.}
\label{fig_cpsi_spectrum}
\end{figure}

Figure~\ref{fig_cpsi_spectrum} demonstrates that the contribution from Born-corrections to the lensing potential spectrum $C^{\psi \psi}_\ell$ is comparable with the additional amplitude caused by nonlinear growth of the large scale structure between us and the CMB. One clearly sees that Born-corrections amplify non-linear contributions to the spectrum because they are proportional to squares of the spectrum. While Born-corrections are negligible up to multipoles of $\ell\simeq3000$ when computed for linear structures, they already cause deviations in $C^{\psi \psi}_\ell$ well below $\ell\simeq 1000$ in the case of nonlinear structures. The identical argumentation applies to $C^{\Omega \Omega}_\ell$ with the only difference that Born-approximated lensing does not show any rotation irrespective of the structure being linear or not.

The difference to the negligible contributions in cosmic shear surveys with typical line of sight integrations up to $z \sim 0.9$ has two main reasons: $(i)$ The line of sight integrations involved in all corrections grow with the distance and effects like lens-lens coupling add up on long paths, and $(ii)$ under the flat-sky Limber equation, the power spectrum is evaluated at $P(k = \ell/\chi)$, so for fixed multipoles we are probing smaller wavenumbers $k$ with larger distance where the amplitude of the CDM spectrum is higher.

% ---  ---%

\begin{figure}
\begin{center}
\resizebox{0.95\hsize}{!}{\includegraphics{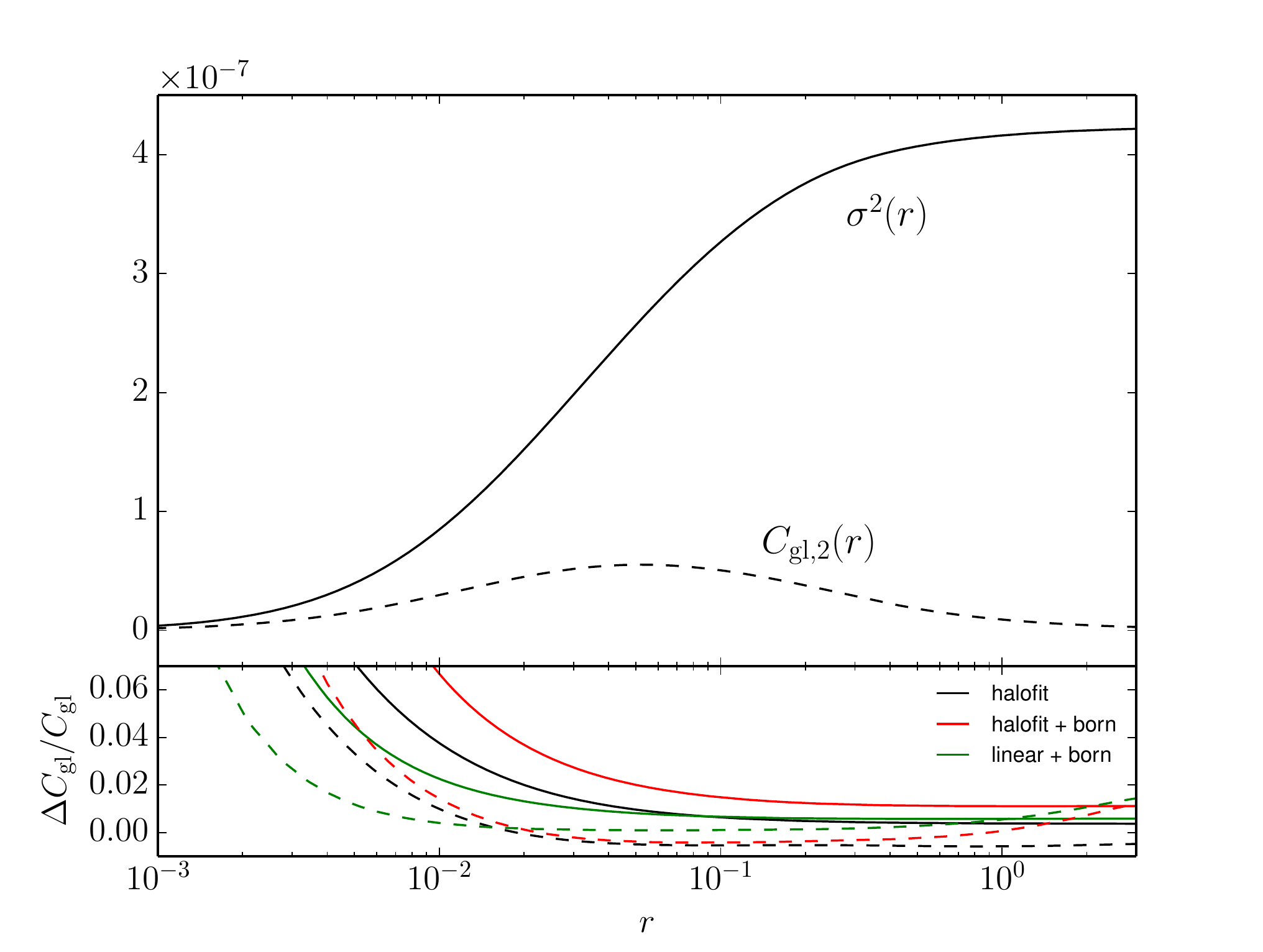}}
\end{center}
\caption{Lensing correlations $\sigma^2(r) \equiv C_\mathrm{gl}(0) - C_\mathrm{gl}(r)$ and $C_{\mathrm{gl},2}(r)$ (dashed) as a function of the angular separation on the sky $r$ in radian. Relative changes introduced by nonlinear structure growth, Born-corrections or both are shown in the lower panel.}
\label{fig_Cgl_combined}
\end{figure}

% ---  ---%
\subsection{CMB temperature}
Gravitational lensing remaps the CMB fields by the deflection angle $\balpha = \delta \bmath x / \chi$, so we observe e.g. the temperature at a lensed position
\begin{equation}
\tilde{\Theta}(\bmath x) = \Theta(\bmath x + \balpha) \: ,
\end{equation}
where for now we take only the part of the deflection angle that may be expressed by the gradient of the underlying lensing potential $\balpha = \nabla \psi$ into account. Again working in the flat-sky limit we can Fourier transform the temperature by
\begin{equation}
\Theta(\bmath l) = \int \frac{\mathrm d^2 \bmath x}{2 \upi} \: \Theta(\bmath x) \: \mathrm e^{\mathrm i \bmath l \cdot \bmath x} \: .
\end{equation}
The deflection is a local effect, so we apply it to the correlation function before transforming back to the power spectrum as done by \cite{Lewis}. One then arrives at the lensed correlation function
\begin{equation}\label{eq_lensed_temp_correlation}
\tilde{\xi}(r) 
= \left\langle \tilde{\Theta}(\bmath{x}) \: \tilde{\Theta}(\bmath{x}^\prime) \right\rangle = \int \frac{\mathrm{d}^2 \bmath{l}}{(2 \upi)^2} C^{\Theta \Theta}_\ell \: \mathrm e^{\mathrm i \bmath{l} \cdot \bmath{r}} \left\langle \mathrm e^{\mathrm i \bmath{l} \cdot (\balpha - \balpha^\prime)} \right\rangle \: ,
\end{equation}
where we introduced $r = | \bmath x - \bmath x^\prime | $ and neglected the integrated Sachs-Wolfe effect which introduces a small correlation between the CMB and the lensing potential.
\subsection{Deflection angle correlations}
In linear theory, the lensing potential obeys Gaussian statistics and inherits this property to the deflection angle. The remaining expectation value in eq.~\ref{eq_lensed_temp_correlation} can then be obtained by using $\langle \mathrm e^{\mathrm i y} \rangle = \mathrm e^{-\langle y^2 \rangle /2}$. This leads to the result \citep{1996ApJ...463....1S}
\begin{equation}\label{eq_corr}
\begin{split}
\left\langle \mathrm e^{\mathrm i \bmath l \cdot (\balpha - \balpha^\prime)} \right\rangle &= \exp \left( -\frac{1}{2} \left\langle [ \bmath l \cdot (\balpha - \balpha^\prime)]^2 \right\rangle \right) \\
&= \exp \left( - \frac{\ell^2}{2} \left[ \sigma^2(r) + \cos 2 \varphi \: C_{\mathrm{gl},2}(r) \right] \right) \: ,
\end{split}
\end{equation}
where $\varphi$ denotes the angle between $\bmath l$ and $\bmath x$ and we introduced the variance of the lensing excursion angle $\sigma^2(r) = \frac{1}{2} \langle (\balpha - \balpha^\prime)^2 \rangle = C_\mathrm{gl}(0) - C_\mathrm{gl}(r)$. The functions $C_\mathrm{gl}$ and $C_{\mathrm{gl},2}$ are given in terms of the lensing potential power spectrum $C^{\psi \psi}_\ell$
\begin{equation}\label{eq_Cgl}
\begin{split}
C_\mathrm{gl}(r) &= \frac{1}{2 \upi} \int \mathrm d \ell \: \ell^3 C_\ell^{\psi \psi} J_0(\ell r) \\
C_{\mathrm{gl},2}(r) &= \frac{1}{2 \upi} \int \mathrm d \ell \: \ell^3 C_\ell^{\psi \psi} J_2(\ell r) \: ,
\end{split}
\end{equation}
and we will model the nonlinear case in the same way by assuming Gaussian statistics while accounting for the additional amplitude of the density power spectrum as discussed in Sect. \ref{sect_gauss}.

The functions $C_\mathrm{gl}$ and $C_{\mathrm{gl},2}$ encode all lensing effects on the spectra. They are shown in Fig.~\ref{fig_Cgl_combined} including nonlinear structure growth and the higher order Born-corrections.

For the numerical implementation, instead of using the computationally demanding expression in eq.~\ref{eq_corr}, we expand it to second order in the off-diagonal part $C_{\mathrm{gl},2}$ of the correlator $\langle \alpha_i \alpha_j \rangle$ as done in \cite{lewis2}. Note that this expansion does not require the deflection itself to be small.

% ---  ---%
\subsection{CMB polarisation}
The CMB polarisation is treated in a very similar way. We start from the trace-free spin-2 polarisation tensor $P = Q + \mathrm i U$ with the Stokes parameters $Q$ and $U$. In the flat-sky limit, the spin-weighted spherical harmonics reduce to $_{\pm2}Y_{\ell m} \to - \mathrm e^{\pm 2 \mathrm i \varphi_\ell} \mathrm e^{\mathrm i \bmath l \cdot \bmath x}$, where $\varphi_\ell$ again denotes the angle between $\bmath l$ and $\bmath x$. Therefore the expansion of $P(\bmath x)$ into the parity eigenstates $E$ and $B$ reads
\begin{equation}
P(\bmath x) = - \int \frac{\mathrm d^2 \bmath l}{2 \upi} \: [ E(\bmath l) + \mathrm i B(\bmath l) ] \mathrm e^{ 2 \mathrm i \varphi_\ell} \mathrm e^{\mathrm i \bmath l \cdot \bmath x} \: .
\end{equation}
In addition, we align all correlation functions with the physical coordinate system along $\bmath r$ by another rotation with the angle $\varphi_r = \hat{\bmath r} \cdot \hat{\bmath x}$. The corresponding correlation functions are then
\begin{equation}\label{eq_polarisation_correlation}
\begin{split}
\xi_+(\bmath r) &\equiv \langle P^\ast_r(\bmath x) \: P_r(\bmath x^\prime) \rangle = \langle P^\ast (\bmath x) \: P(\bmath x^\prime) \rangle \\
\xi_-(\bmath r) &\equiv \langle P_r(\bmath x) \: P_r(\bmath x^\prime) \rangle = \langle \mathrm e^{-4i\varphi_r} P(\bmath x) \: P(\bmath x^\prime) \rangle \\
\xi_\times (\bmath r) &\equiv \langle P_r(\bmath x) \: \Theta(\bmath x^\prime) \rangle = \langle \mathrm e^{-2i\varphi_r} P(\bmath x) \: \Theta(\bmath x^\prime) \rangle \: .
\end{split}
\end{equation}
The calculation of the lensed correlation function $\xi_+$ is completely analogous to the temperature case, whereas for the remaining functions the additional phase factors must be taken into account. Full expressions for the lensed correlation functions used here can be found in \cite{1998PhRvD..58b3003Z} and \cite{lewis2}.

% ---  ---%
\subsection{Lensing by the Curl potential}\label{sect_curl}
\begin{figure}
\begin{center}
\resizebox{0.95\hsize}{!}{\includegraphics{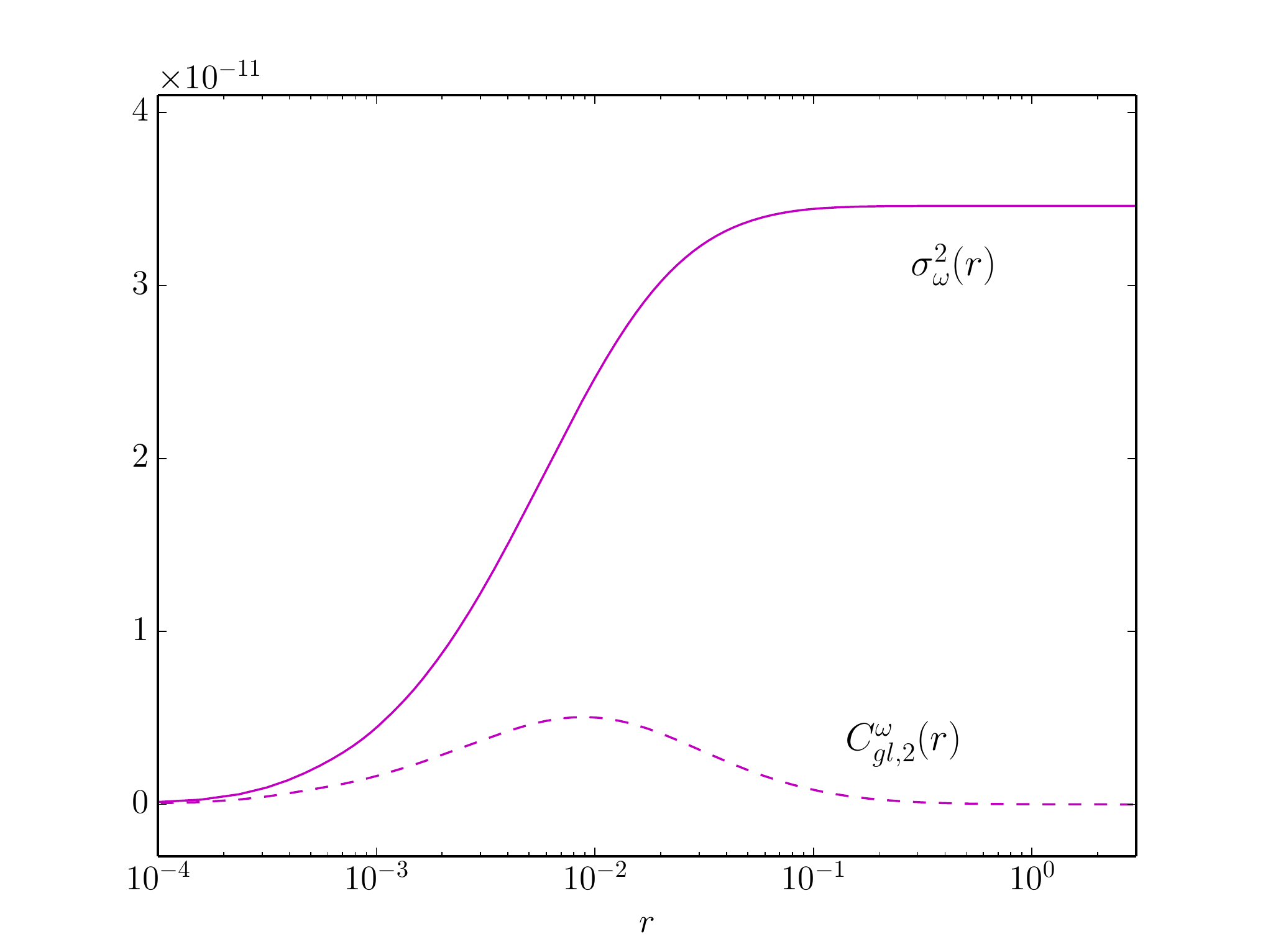}}
\end{center}
\caption{Curl-like lensing correlations $\sigma_\omega^2(r) \equiv C_\mathrm{gl}^\omega(0) - C_\mathrm{gl}^\omega(r) $ and $C_{\mathrm{gl},2}^\omega(r)$ (dashed) as functions of the angular seperation on the sky $r$ in radian. Compared to the functions $\sigma^2$ and $C_{\mathrm{gl},2}$ derived from the lensing potential $\psi$ shown in Fig.~\ref{fig_Cgl_combined} the overall scale is suppressed by four orders of magnitude.}
\label{fig_Cgl_omega}
\end{figure}

After we dealt with the gradient-like part of the lensing deflection angle, now we turn to the curl-like contribution $\boldeta = \nabla \times \Omega$ which only appears if Born-corrections are included. We start in analogy to the previous case by considering the lensed temperature correlation function
\begin{equation}\label{eq_rot_correlator}
\tilde{\xi}(r) 
= \int \frac{\mathrm{d}^2 \bmath{l}}{(2 \upi)^2} C^{\Theta \Theta}_\ell \: \mathrm e^{\mathrm i \bmath{l} \cdot \bmath{r}} \left\langle \: \mathrm e^{\mathrm i \bmath{l} \cdot (\boldeta - \boldeta^\prime)} \right\rangle \: .
\end{equation}
Again we have to evaluate the correlator. The full calculation shown in Appendix \ref{appendix_rotangle} leads to:
\begin{equation}
\left\langle \mathrm e^{\mathrm i \bmath l \cdot (\boldeta - \boldeta^\prime)} \right\rangle 
= \exp \left( - \frac{\ell^2}{2} \left[ \sigma_\omega^2(r) + \cos 2 \varphi \: C^\omega_{\mathrm{gl},2}(r) \right] \right) \: ,
\end{equation}
where we defined $\sigma_\omega^2(r) = C^\omega_\mathrm{gl}(0) - C^\omega_\mathrm{gl}(r)$ and the two functions
\begin{equation}\label{eq_Cgl_curl}
\begin{split}
C_\mathrm{gl}^\omega(r) &= \frac{1}{2 \upi} \int \mathrm d \ell \: \ell^3 C_\ell^{\Omega \Omega} J_0(\ell r) \\
C_{\mathrm{gl},2}^\omega(r) &= - \frac{1}{2 \upi} \int \mathrm d \ell \: \ell^3 C_\ell^{\Omega \Omega} J_2(\ell r) \: .
\end{split}
\end{equation}
The main difference to the expressions in eq.~\ref{eq_Cgl} is the replacement of the power spectrum $C^{\psi \psi}_\ell$ by $C^{\Omega \Omega}_\ell$ and the opposite sign of $C_{\mathrm{gl},2}^\omega$. Note that the exponential growth proportional to $C_{\mathrm{gl},2}^\omega$ does not pose a problem because we find $\sigma_\omega^2 > C_{\mathrm{gl,2}}$ everywhere, so the damping is always dominant. Both functions are shown in Fig.~\ref{fig_Cgl_omega}. They follow a very similar form as their $C^{\psi \psi}_\ell$-counterparts, but the overall scale is four orders of magnitude below the correlation functions computed for the gradient-like lensing potential shown in Fig.~\ref{fig_Cgl_combined}. The effect of image rotations by lensing on the CMB is therefore not expected to play any role in future measurements. Therefore the generation of lensed CMB-maps by the mapping $\tilde{\Theta}(\bmath x) = \Theta(\bmath x + \balpha)$ as computed by \textsc{lenspix} where the deflection angle is derived through the gradient of the lensing potential, $\balpha = \nabla \psi$, remains an excellent approximation.

% ---  ---%
\subsection{Resulting CMB spectra}\label{sec_results}
To compute the Born-corrected CMB spectra, we use $C^{\psi\psi}_\ell$ including all corrections up to $\phi^4$ to calculate the correlation functions $\sigma^2$ and $C_{\mathrm{gl},2}$ and apply the lensing procedure outlined in Sect.~\ref{sect_lensing} to the initial, unlensed temperature and polarisation spectra as taken from \textsc{camb} \citep{camb}. Considering the results presented in Fig.~\ref{fig_Cgl_omega}, we neglect any lensing contributions from the curl potential $\Omega$ as discussed in Sect.~\ref{sect_curl}.

\begin{figure*}
\subfigure{\includegraphics[width=.45\linewidth]{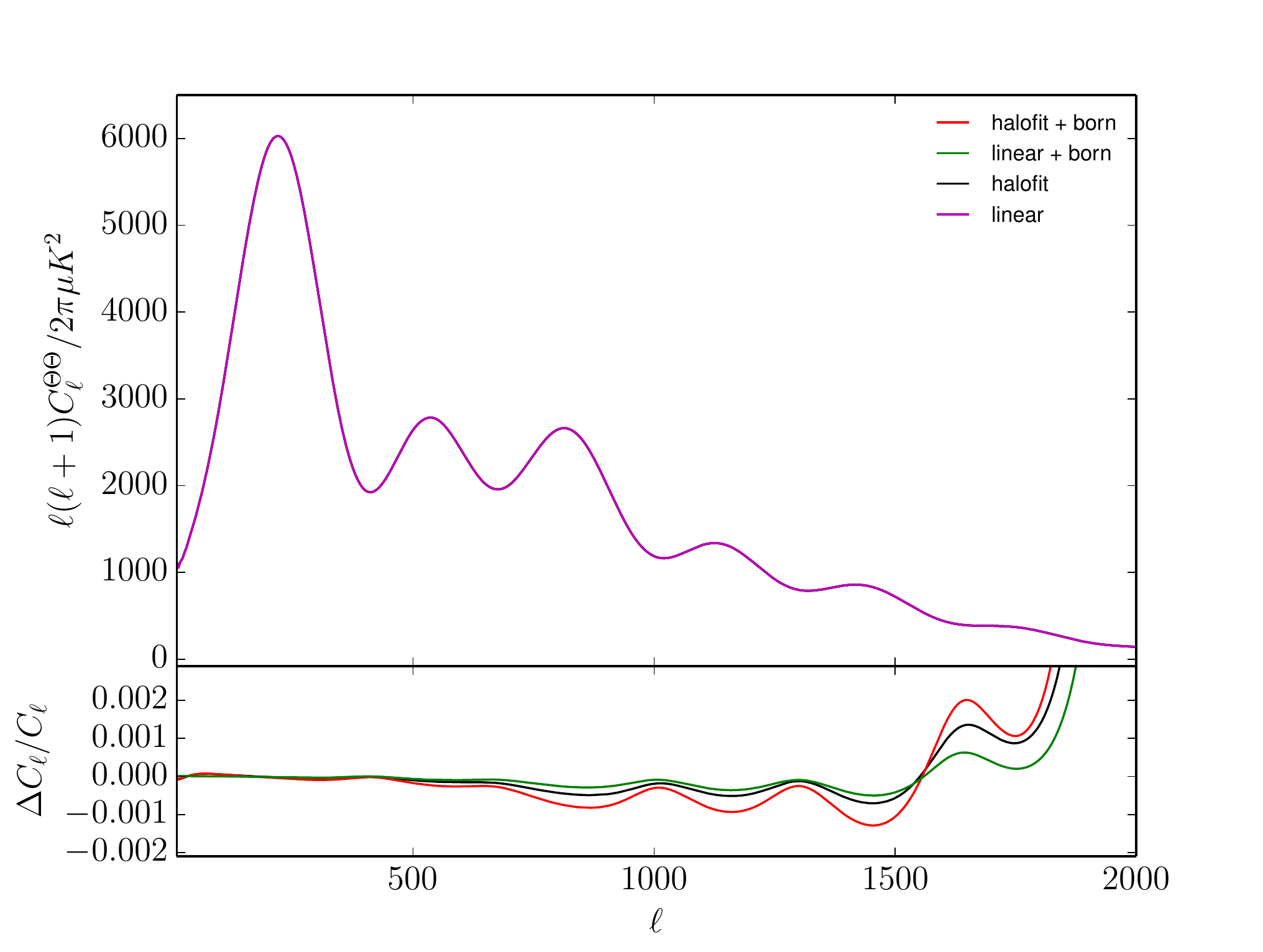}
}
\subfigure{\includegraphics[width=.45\linewidth]{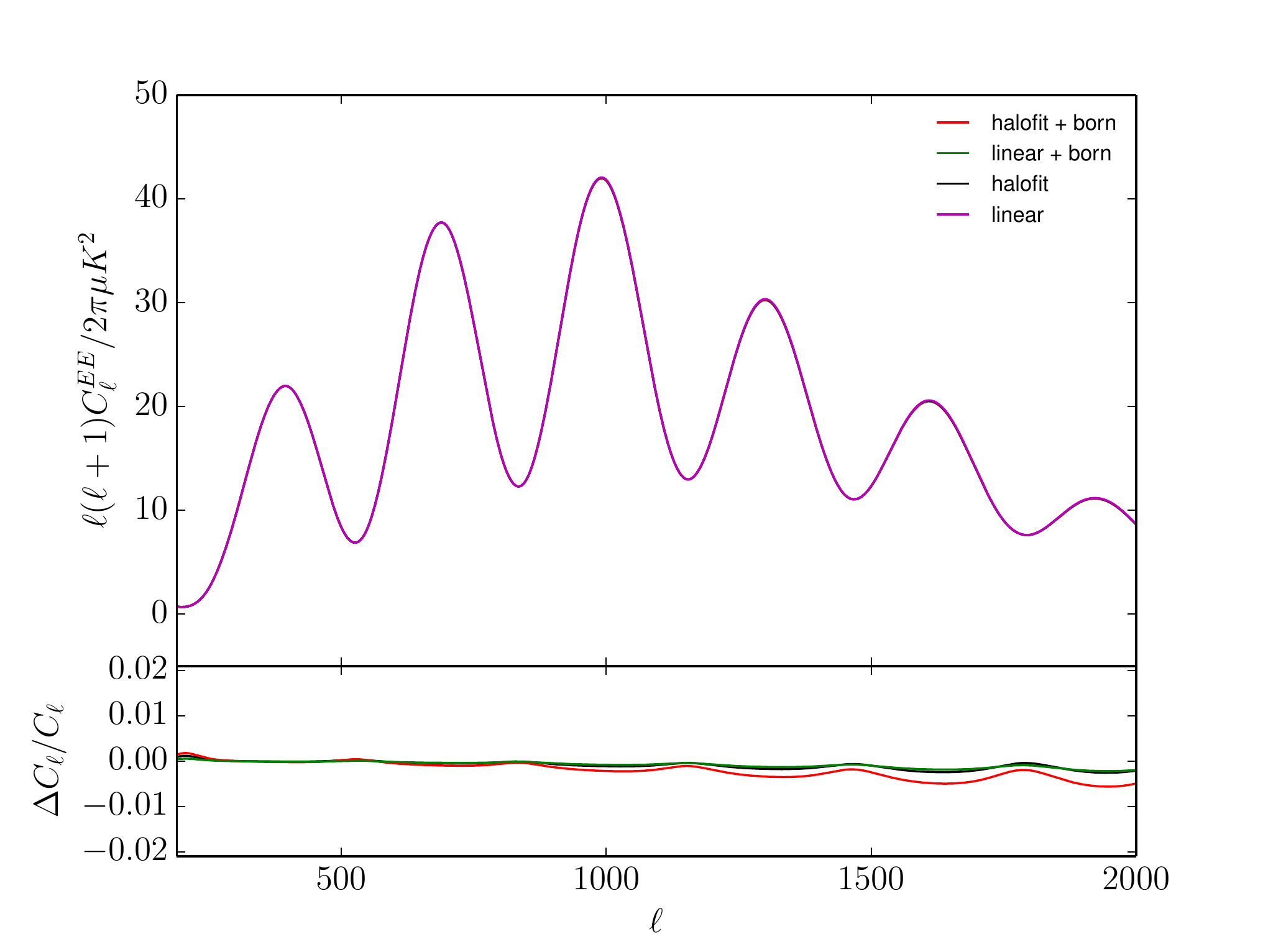}
} \\
\subfigure{\includegraphics[width=.45\linewidth]{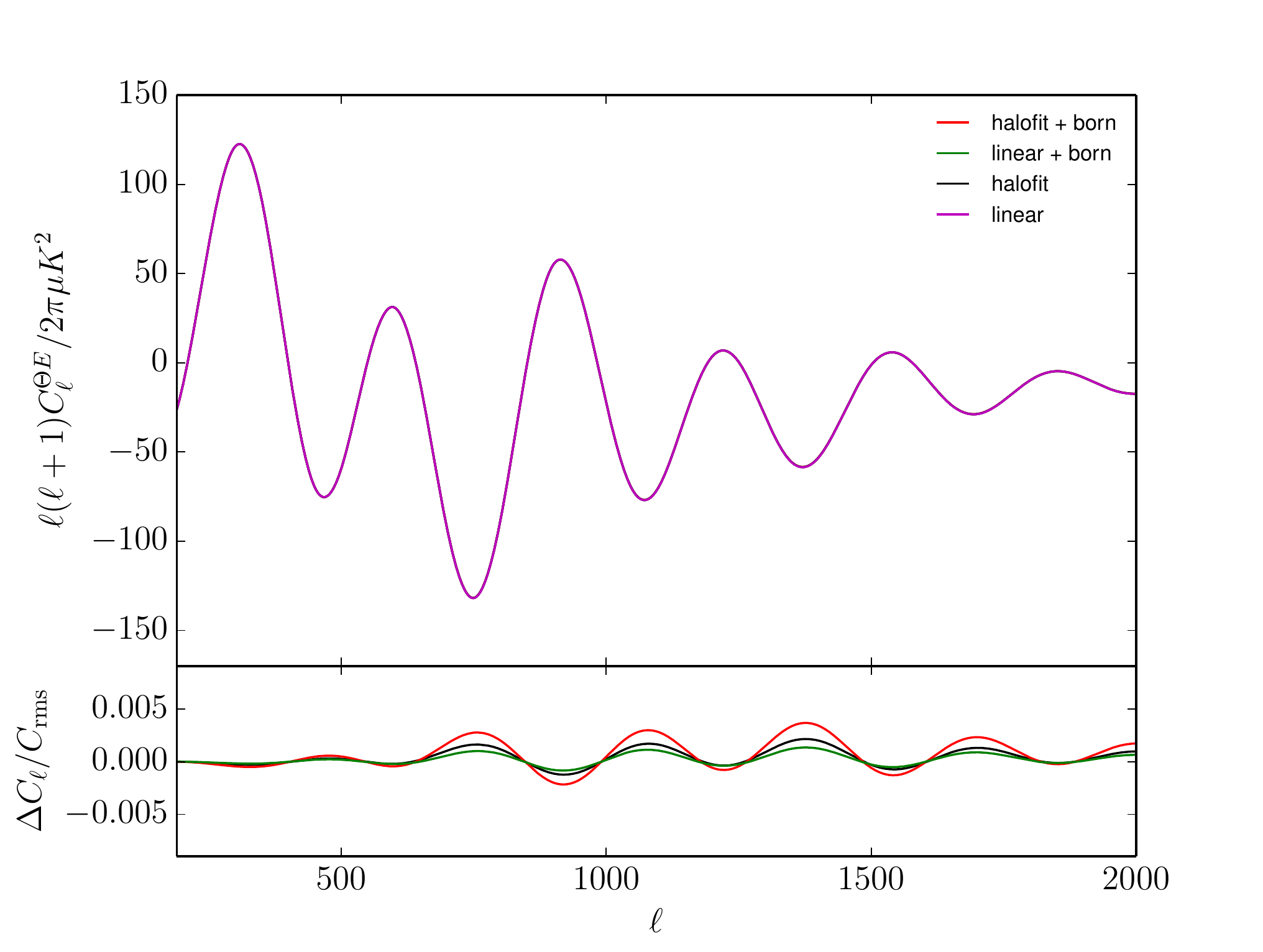}
}
\subfigure{\includegraphics[width=.45\linewidth]{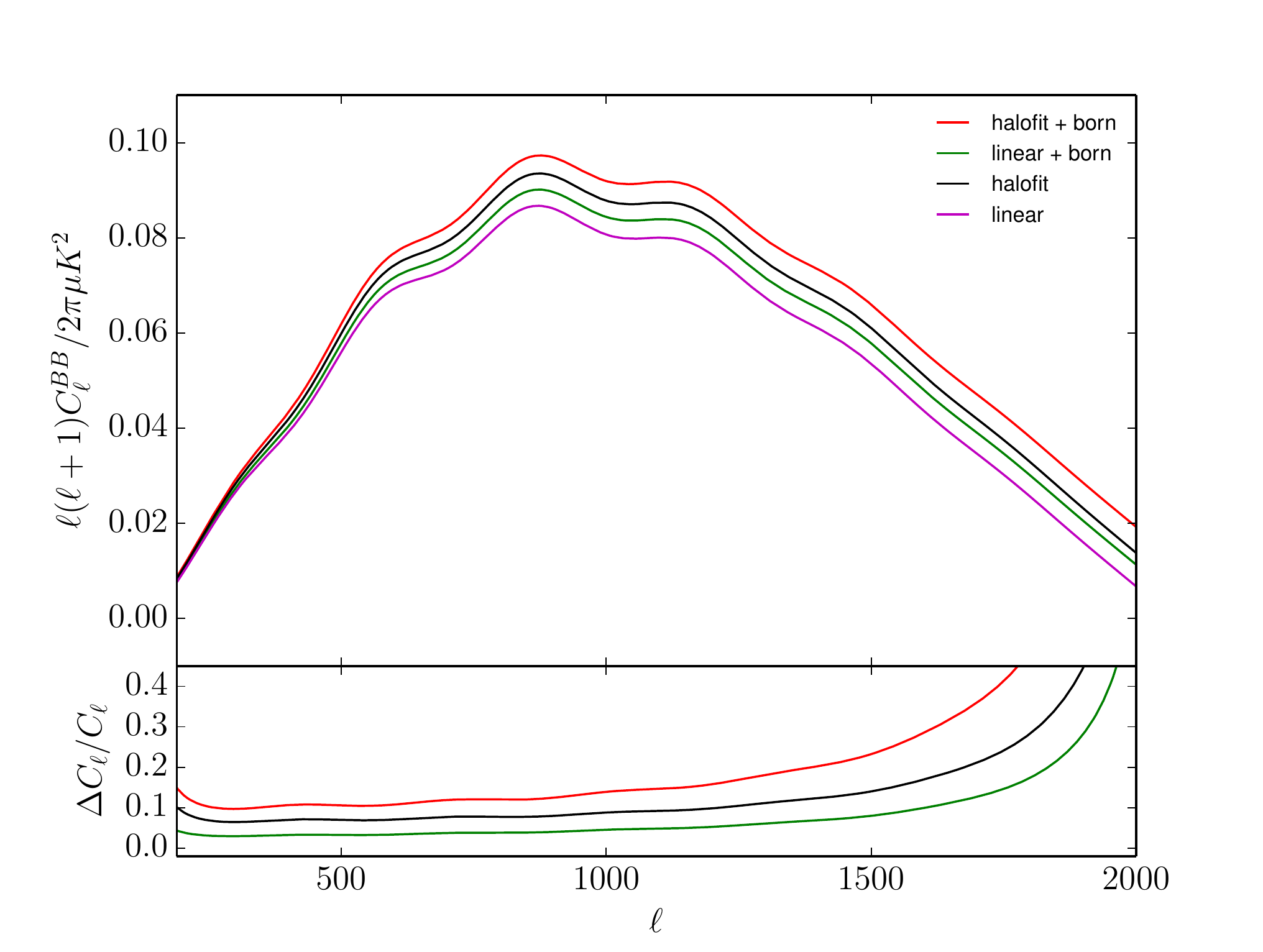}
}\caption{The lensed temperature and polarisation spectra. Shown are the first order lensing results with linear CDM power (violet), nonlinear power (black) and including all Born-corrections in combination with linear (green) and nonlinear power (red). Relative corrections compared to the first-order lensing result with linear CDM spectrum are shown below each plot. Tensor modes are not included, so the unlensed $C^{BB}$ spectrum vanishes. Note that the $C^{\Theta E}$ spectrum is normalised to the rms value to avoid divisions by zero.}
\label{fig_spectra}
\end{figure*}

The broad peak structure of the CMB temperature spectrum is less susceptible to lensing in general as the sharper features of the polarisation, therefore the corrections discussed here are less important when applied to $C^{\Theta \Theta}_\ell$ as illustrated by Fig.~\ref{fig_spectra}. They only lead to relative corrections of $\sim 10^{-4}$ for small scales $\ell > 1000$.

%\begin{figure}
%\begin{center}
%\resizebox{0.95\hsize}{!}{\includegraphics{./figures/TTcombined.pdf}}
%\end{center}
%\caption{The $C^{\Theta\Theta}_\ell$ temperature spectrum calculated including nonlinear structure growth and all second order corrections. Shown are the first order lensing result with linear CDM spectrum (violet), nonlinear CDM spectrum (black) and including Born-corrections in combination with linear (green) and nonlinear (red) power. Relative corrections to the $C^{\Theta\Theta}_\ell$ spectrum compared to the first-order lensing result with linear CDM spectrum are shown below.}
%\label{fig_TTrel}
%\end{figure}

The change is more pronounced in the polarisation. As shown in Fig.~\ref{fig_spectra}, the $E$-mode spectrum $C^{EE}_\ell$ is modified by $\sim 10^{-2}$ for large multipoles. Consequently the corrections to the cross-spectrum between temperature and $E$-mode polarisation fall in between the previous cases, as shown in Fig.~\ref{fig_spectra}, giving rise to relative deviations of the order $10^{-3}$. There is a general trend that the Born-corrections affect polarisation spectra more strongly than temperature spectra.

%\begin{figure}
%\begin{center}
%\resizebox{0.95\hsize}{!}{\includegraphics{./figures/TEcombined.pdf}}
%\end{center}
%\caption{The $C^{\Theta E}_\ell$ polarisation spectrum calculated including nonlinear structure growth and all second order corrections. Shown are the first order lensing result with linear CDM spectrum (violet), nonlinear CDM spectrum (black) and including Born-corrections in combination with linear (green) and nonlinear (red) power. Relative corrections to the $C^{\Theta E}_\ell$ polarisation compared to the first-order lensing result with linear CDM spectrum are shown below. Note that the relative deviation is normalised to the rms value.}
%\label{fig_TErel}
%\end{figure}

%\begin{figure}
%\begin{center}
%\resizebox{0.95\hsize}{!}{\includegraphics{./figures/EEcombined.pdf}}
%\end{center}
%\caption{The primordial $C^{EE}_\ell$ polarisation spectrum calculated by \textsc{camb} (black) and the lensed results using a nonlinear power spectrum in combination with Born-corrections (red) and with the first order result (green). Relative corrections to the $C^{EE}_\ell$ polarisation compared to the first-order result with linear CDM spectrum are shown on the bottom.}
%\label{fig_EErel}
%\end{figure}

Finally, the $B$-mode spectrum $C^{BB}_\ell$ is most strongly affected, with changes of typically 10 per cent relative to the Born-approximated result, already at comparatively low multipoles. This is related to the fact that lensing is the strongest $B$-mode generating effect in the absence of foregrounds and of tensor modes \citep{BICEP,planckxxx}, such that the Born-corrections do not compete with a large intrinsic polarisation signal.

%\begin{figure}
%\begin{center}
%\resizebox{0.95\hsize}{!}{\includegraphics{./figures/BBcombined.pdf}}
%\end{center}
%\caption{The $C^{BB}_\ell$ polarisation spectrum calculated including nonlinear structure growth and all second order corrections. Tensor perturbations are neglected so the unlensed spectrum is zero. Shown are the first order lensing result with linear CDM spectrum (violet), nonlinear CDM spectrum (black) and including Born-corrections in combination with linear (green) and nonlinear (red) power. Relative corrections to the $C^{BB}_\ell$ polarisation compared to the first-order lensing result with linear CDM spectrum are shown below.}
%\label{fig_BBrel}
%\end{figure}

In summary, Born-corrections affect the lensing signal on small angular scales and are similar in magnitude compared to nonlinear corrections of the deflection field. Their influence is larger for polarisation compared to temperature anisotropies, and they are most notable in the $B$-mode signal.

% --- section: summary --- %
\section{Summary}\label{sect_summary}
Subject of this paper were Born-corrections in the lensing of the CMB temperature and polarisation anisotropies. We pursued an analytical technique and applied a perturbative solution to the lens equation up to cubic terms in the gravitational potential to construct the next-to-leading-order corrections $\propto \phi^4$ to the weak lensing power spectrum. From the physical point of view this corresponds to replacing the fiducial straight ray along which the lensing deflections are summed by an integration path closer to the true photon geodesic for collecting the lensing deflections. The point at which the lensing deflections are evaluated is related to the corresponding location on a fiducial straight ray by a Taylor expansion, where the distance between these two points follows from cumulative lensing up to that point. This leads to a mode-coupling effect in the deflection field due to the correlations along the line of sight, which source Born-corrections.

Computing corrections to the spectrum of the lensing potential~$C^{\psi \psi}_\ell$ shows almost perfect agreement with the Born-approximated spectrum up to multipoles of $\ell\simeq1000$, and growing differences with increasing multipoles. Already at larger angular scales lensing from nonlinear structures starts to increase the variance of the deflection field, such that the combined effect, where Born-corrections are computed with the nonlinear CDM-spectrum, dominates over the Born-approximated and linear spectrum by almost an order of magnitude at high multipoles.

While this sounds dramatic, it should be kept in mind that the correlation functions of the lensing deflection angle, which are the relevant quantities in transforming the CMB temperature and polarisation spectra, result from a weighted integration over the spectrum of the lensing potential with Bessel functions $J_0(\ell r)$ and $J_2(\ell r)$ cutting off contributions from the high-$\ell$ part of $C^{\psi \psi}_\ell$. Consequently, the remaining relative deviations in the angular correlation functions of the deflection angle amount typically to a few percent. 

Repeating the calculation for the rotational part of the deflection field, which can not be derived from a scalar lensing potential, shows very small angular correlation functions which can be safely neglected: From this point of view the approach of the \textsc{lenspix} code for computing lensed realisations of CMB maps by deriving lensing displacements $\balpha = \nabla \psi$ is perfectly justified. Born-corrections would amount to typically a few percent of the the lensing deflection, which is safe to use in a linear mapping.

The additional amplitude in the lensing potential spectrum $C^{\psi\psi}_\ell$ for multipoles $\ell > 1000$ is larger than the contribution of nonlinear structure growth as modeled by \textsc{halofit}. There is a considerable effort made to accurately model the latter \citep{Carbone1,Carbone2,Carbone3}, in particular for non-standard cosmologies. Born-corrections are amplified by nonlinear corrections to the CDM-spectra, because of the convolution of the spectra with themselves, which implies that both corrections should be linked to each other.

Our results suggest that Born-corrections should be taken into account to calculate $E$-mode polarisation spectra to the per cent level while the effect on $B$-modes is even larger. Because the $B$-mode spectrum is sensitive to a wide range of multipoles, Born-corrections introduce relative changes of order 10 per cent on all scales. CMB-spectra involving the temperature anisotropy are almost unaffected by Born-corrections, which would offer the possibility of consistency checks between the spectra $C^{EE}_\ell$ and $C^{BB}_\ell$ on one side and $C^{\Theta\Theta}_\ell$ and $C^{\Theta E}_\ell$ on the other.

Our results are in contrast to those by \citet{2014arXiv1409.7680C}, who computed multiple lensing on a simulation. In their setup they propagated CMB temperature and polarisation maps with \textsc{lenspix} through a nested set of deflection angle maps which had been derived from numerical simulations of cosmic structures, and did not find deviations due to multiple lensing. Commenting on this we are uncertain if Born-corrections are correctly evaluated by their numerical approach: While it is certain that \citet{Carbone3} have a very accurate description of growing cosmic structures including halo formation and lensing on the steep gradients close to halo centres on small scales similar to \citet{2005MNRAS.356..829H}, it seems that the construction of shells from reoriented tesselated simulation boxes and collapsing them onto lens planes would destroy correlations in the lensing deflection along the line of sight, which is necessary for Born-corrections to arise. We may roughly estimate the radial resolution needed for achieving convergence between perturbation theory and simulation following \citet{Lewis}: Photons traverse a (comoving) distance of 14~Gpc on their way from the last scattering surface to today's observer. The typical size of a lens can be approximated by the comoving BAO scale of the dark matter power spectrum $r_s \sim 150$~Mpc, leading to a typical diameter of $\sim 300$~Mpc. Accordingly, CMB photons undergo approximately 50 deflections. Thus, a comparable number of correlated radial shells should be able to resolve Born-corrections due to multiple lensing. Such high resolution simulations, which cover sufficiently large length scales, are challenging but first successful attempts have been made, for example by the Horizon~$4\upi$ simulation \citep{2009A&A...497..335T}.

Nonlinear structures generate non-Gaussian statistical properties of any observable. Surprisingly, the effect of nonlinear structures is mainly contained in an increased variance of the deflection field, in the limit of lensing by a smooth density field. Expanding the characteristic function of the lensing deflection angle distribution in a series shows only small changes due to higher-order moments \citep{PhilippGauss}. This is valid on scales up to an arcminute, below which lensing on individual haloes adds slowly decreasing wings to the distribution, as shown by \citet{2005MNRAS.356..829H}. Numerical simulations, on the other side, contain a correct description of haloes and of nonlinear structures for a large range of cosmological models \citep{Carbone1,Carbone2,Carbone3}, which is difficult to reach by perturbation theory.

Summarising we would like to point out that Born-corrections are relevant for CMB-lensing at a similar level as nonlinear corrections to the CDM spectrum. While these are not necessary for current observations of CMB-lensing,  they will become relevant in the future.

% --- section: acknowledgements --- %
\section*{Acknowledgements}
We would like to thank C. Carbone for useful discussions on simulations of CMB lensing.

% --- section: bibliography --- %
\bibliography{bibtex/aamnem,bibtex/references}
\bibliographystyle{mn2e}

% --- appendix --- %
\appendix

% --- section: Born --- %
\section{Born correction terms}\label{appendix_born}
We quickly give an overview of all source terms up to order $\phi^3$ appearing in addition to the first order result $S^{(1)}_{ij} = \phi_{ij}$ when dropping the Born approximation as described in Sect. \ref{subsect_born}. \\
Born-corrections:
\begin{equation}
\begin{split}
S^{(2)}_{ij, B} &= -\phi_{ijk}(\chi) \: 2 \int_0^\chi \mathrm d \chi^\prime \: W(\chi^\prime, \chi) \: \frac{\chi}{\chi^\prime} \: \phi_k(\chi^\prime) \\
S^{(3)}_{ij, B} &= \frac{1}{2} \phi_{ijkm}(\chi) \: 2 \int_0^\chi \mathrm d \chi^\prime \: W(\chi^\prime, \chi) \: \frac{\chi}{\chi^\prime} \: \phi_k(\chi^\prime) \\
& \phantom{=} \: \times 2 \int_0^{\chi^{\prime \prime} } \mathrm d \chi^{\prime \prime} W(\chi^{\prime \prime}, \chi ) \: \frac{\chi}{\chi^{\prime \prime} } \: \phi_m(\chi^{\prime \prime} ) \\
S^{(3)}_{ij, BX} &= \phi_{ijm} (\chi) \: 2 \int_0^\chi \mathrm d \chip \: W(\chip, \chi) \: \frac{\chi}{\chip} \: \phi_{mk}(\chip) \\
& \phantom{=} \: \times 2 \int_0^\chip \mathrm d \chipp \: W(\chipp, \chip) \: \frac{\chip}{\chipp} \: \phi_k(\chipp) \: .
\end{split}
\end{equation}
Lens-lens-corrections:
\begin{equation}
\begin{split}
S^{(2)}_{ij, L} &= \phi_{im}(\chi) \: 2 \int_0^\chi \mathrm d \chip W(\chip, \chi) \: \phi_{mj}(\chip) \\
S^{(3)}_{ij, L} &= \phi_{im}(\chi) \: 2 \int_0^\chi \mathrm d \chip W(\chip, \chi) \: \phi_{mk}(\chip) \\
& \phantom{=} \: \times 2 \int_0^\chip \mathrm d \chipp \: W(\chipp, \chip) \: \phi_{kj}(\chipp) \: .
\end{split}
\end{equation}
Mixed terms:
\begin{equation}
\begin{split}
 S_{ij, \mathrm{BL}}^{(3)} = & \: \phi_{ikm}(\chi) \: 2 \int^\chi \mathrm{d} \chi^\prime \: W(\chi^\prime, \chi) \: \frac{\chi}{\chi^\prime} \: \phi_k (\chi^\prime)  \\ 
 & \times 2 \int^{\chi^\prime} \mathrm{d} \chi^{\prime \prime} \: W(\chi^{\prime \prime}, \chi^\prime) \: \phi_{mj} (\chi^{\prime \prime} )	\: .
\end{split}
\end{equation}
Under the assumption of Gaussianity, the perturbative corrections to the angular spectrum of order $\phi^4$ can then be constructed from the sum of all combinations $\langle S^{(2)} S^{(2)} \rangle$ and $\langle S^{(3)} S^{(1)} \rangle$ as explained in Sect. \ref{subsect_spectra}.

% --- section: rotational part --- %
\section{Angular correlation function of the rotation angle}\label{appendix_rotangle}
To compute the expectation value in eq.~\ref{eq_rot_correlator}, we have to evaluate the correlator $\langle \eta_i \eta_j \rangle$. We split it up into a diagonal part $C^\omega_\mathrm{gl}$
\begin{equation}
C_\mathrm{gl}^\omega (r) = \langle \boldeta \cdot \boldeta^\prime \rangle = \int \frac{\mathrm d^2 \bmath l}{2 \upi} \: \epsilon_{ij} \ell_{j} \epsilon_{ik} \ell_k \: C^{\Omega \Omega}_\ell \mathrm e^{\mathrm i \bmath l \cdot \bmath r}
\end{equation}
and using $\epsilon_{ij} \epsilon_{ik} = \delta_{jk}$ we arrive at
\begin{equation}
C_\mathrm{gl}^\omega (r) = \frac{1}{2 \upi} \int \mathrm d \ell \: \ell^3 C_\ell^{\Omega \Omega} J_0(\ell r) \: ,
\end{equation}
which is identical to the expression in eq.~\ref{eq_Cgl} with the potential spectrum $C^{\psi \psi}_\ell$ replaced by $C^{\Omega \Omega}_\ell$. For the remaining trace-free part we get:
\begin{equation}
\begin{split}
\langle \eta_i \eta_j \rangle \hat r_i \hat r_j &= \frac{1}{2} \big[C_\mathrm{gl}^\omega - C_{\mathrm{gl},2}^\omega \big] \\
&= \int \frac{\mathrm d^2 \bmath l}{(2 \upi)^2} \epsilon_{ik} \ell_k \hat{r}_i \: \epsilon_{jm} \ell_m \hat r_j \: C^{\Omega \Omega}_\ell \: \mathrm e^{i \bmath l \cdot \bmath r} \\
&= \int \frac{\mathrm d^2 \bmath l}{(2 \upi)^2} \: \ell^2 \sin^2 \varphi \: C^{\Omega \Omega}_\ell \: \mathrm e^{i \ell r \cos \varphi} \\
&= \int \frac{\mathrm d \ell}{(2 \upi)^2} \: \ell^3 \: C_\ell^{\Omega \Omega} \int \mathrm d \varphi \: \frac{1}{2} (1- \cos 2 \varphi) \: \mathrm e^{i \ell r \cos \varphi} \\
&= \int \frac{\mathrm d \ell}{4 \upi} \: \ell^3 C_\ell^{\Omega \Omega} \: \big[ J_0(\ell r) - J_2(\ell r) \big] \: .
\end{split}
\end{equation}
For the third line we used that the $\epsilon$-tensor in two dimensions rotates a vector by $\upi/2$ counterclockwise. So if $\varphi$ denotes the angle between~$\bmath r$ and~$\bmath l$, we get $\epsilon_{ij} \ell_i \hat r_j = \ell \cos(\varphi - \frac{\upi}{2}) = \ell \sin \varphi$. This yields the off-diagonal correlation function
\begin{equation}\label{C_omega2}
C_{\mathrm{gl},2}^\omega = - \int \frac{\mathrm d \ell}{2 \upi} \ell^3 C_\ell^{\Omega \Omega} J_2(\ell r) \: .
\end{equation}

\bsp

\label{lastpage}

\end{document}